\documentclass[aps,prb,showpacs,twocolumn,preprintnumbers,amsmath,amssymb,amsfonts,floatfix]{revtex4}

\usepackage{graphicx,color,rotating,textcomp} 
\usepackage{dcolumn} 
\usepackage{bm} 
\newcommand{\sgn}[1]{\text{sgn}\left(#1\right)}
\newcommand{\rv}{\vec{r}}
\newcommand{\rvp}{\vec{r}^{\;\prime}}

\newcommand{\eq}[1]{(\ref{#1})}

\begin{document}

\title{Conductance footprints of impurity scattering in graphene nanoribbons}

\author{Anders~Bergvall}
\author{Tomas~L\"ofwander}
\affiliation{Department of Microtechnology and Nanoscience - MC2,
Chalmers University of Technology, SE-412 96 G\"oteborg, Sweden}

\date{\today}

\begin{abstract}

We report a detailed analytic investigation of the interplay between size quantization
and local scattering centers in armchair graphene nanoribbons, as seen in the conductance.
The scattering property of a local scattering center is dependent on if it is located on one sublattice
(A-site impurity) or both (impurity situated at neighboring carbon atoms, A-B-site impurity).
The A-site impurity scatters in a similar fashion as a localized impurity in a one-dimensional
channel made from a two-dimensional electron gas. On the other hand, the A-B-site impurity
includes multiple scattering involving A- to B-sublattice scattering, which knows about
the chirality of the Dirac electrons and heavily influence the conductance.
For A-site impurities, an intricate interplay between evanescent waves at the impurity
and the propagating modes contributing to the conductance, leads to scattering resonances
that generate either dips in the conductance or render the impurity completely transparent.
The latter occurs at subband bottom energies $E_{n+1}$, where the wave vector of mode $n+1$ turns to zero.
The conductance $G=nG_0$ ($G_0$ the conductance quantum) of defect free graphene therefore remains
at these energies after adding one A-site impurity.
This is analagous to the case of a scattering center in a quantum channel made from a two-dimensional electron gas.
The conductance dips occur at energies $\Delta E$ from the conductance 
steps and their location depend directly on the impurity strength $\gamma$.
In particular, for repulsive impurities ($\gamma>0$) the dips occur for hole-doping,
while for attractive impurities ($\gamma<0$) for electron doping.
For an A-B-site impurity, the A- to B-sublattice scattering interferes with the transmission resonance at the energies
of the subband bottoms and the impurity is never transparent and the conductance steps of defect free graphene
ribbons are always lost.
We derive a generalized Fisher-Lee formula for graphene leads that holds
for arbitrary scattering region and arbitrary number of leads.

\end{abstract}


\maketitle

%
%

\section{Introduction}

A great challenge in graphene technology is the fabrication of nanostructures free from disorder and
impurities\cite{CastroNetoReview,PeresReview,NovoselovRoadmap} (such as defects, 
adatoms, vacancies and edge disturbances). With a lot of research put into the field, the graphene community has gotten closer to 
the goal of producing such atomically clean structures. Fabrication methods include both top-down approaches (such as scanning 
probe methods\cite{ConnollyReview,DeshpandeReview}, nano-litography,\cite{HanPRL2007} directional etching using nanoparticles\cite{CamposNanoLett2009} or rearranging atoms using TEM\cite{GiritScience2009}) and bottom-up approches 
(such as chemical synthesis\cite{CaiNature2010} or the unzipping of carbon nanotubes\cite{LiScience2008}).
These have all proven to be useful, but so far not perfect.\cite{BanhartReview}

Experiments on one type of such nanostructures, graphene nanoribbons (GNRs), reveal that they are not ideal, and much work, 
theoretical and experimental, has been aimed at explaining how the transport properties of GNRs are effected by the disorder and 
impurities normally present in the ribbons. Typical attributes of non-ideal ribbons, such as absence of conductance quantization 
and the development of a conductance gap, has thus been given much attention,
see the reviews in Ref.~\onlinecite{RazaCollection,KatsnelsonBook,RocheBook}.

In this paper, we make an analytical study of graphene nanoribbons with perfect armchair edges (ANGRs). In particular, we examine 
in detail how the interplay between size-quantization and localized scattering centers changes the transmission properties of the ribbons. We 
base our analytical work on a Fisher-Lee type relation\cite{FisherLee} between the single-electron propagators and the current-transmission 
amplitudes. As seen in many numerical studies,\cite{RazaCollection,KatsnelsonBook,RocheBook}
we find that the presence of a localized scattering center causes characteristic 
dips in the conductance through the system when the Fermi energy approaches the bottom of a higher energy subband.
The origin of this behavior is the build-up of evanescent waves and quasi-bound states around the 
scatterer that causes resonant backscattering of the propagating waves.
We also show that, for A-site impurities (see below), the backscattering 
vanishes completely and the transmission recovers its original value when the Fermi energy exactly touches the higher energy subband.
This is analagous to what has been found in previous works on one-dimensional channels
made from two-dimensional electron gases.\cite{Bagwell1,Bagwell2}
For A-B-site impurities, the resonances are heavily influenced by the A- to B-site scattering events. This is unique to graphene
and we will highlight its consequences.

\section{Impurity scattering including quantum confinement}\label{ch:scattering}

The problem of scattering against a localized defect in an otherwize perfect 2D sheet of graphene has been studied in several works,
see for instance the review in Ref.~\onlinecite{WehlingReview}
In particular, it has been shown that impurity resonances (quasibound) states can be formed.
If the potential strength $\gamma$ of the localized impurity is very large,
the resonance can approach the Dirac point according to the approximate formula $E_{imp}\approx D^2/(2\gamma \ln |D/(2\gamma)|)$,
where $D$ is a high-energy cut-off beyond which the Dirac approximation is invalid ($\sim 2|t|$ in graphene, where $t$ the nearest neighbor
hopping amplitude). In the presence of a random distribution of such impurities, an impurity band may form which at low temperatures
and energies can dominate transport properties.\cite{GusyninPRL2005,PeresPRB2006,LofwanderPRB2007}

\begin{figure}[t]
\includegraphics[width=\columnwidth]{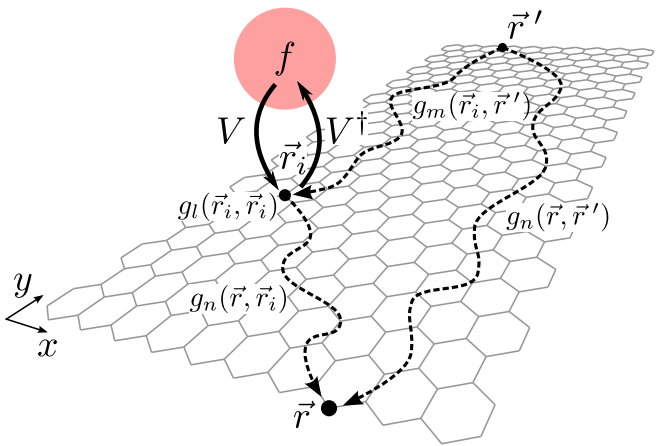}
\caption{The process of electron scattering in an armchair graphene nanoribbon. The electron may propagate directly from $\rvp$ to 
$\rv$ in transverse mode number $n$, or start in mode $m$ at $\rvp$ and take a detour to interact
with the impurity at $\rv_i$ and end up in mode $n$ at $\rv$ after multiple scattering events involving modes $l$ at $\rv_i$.
The propagator from $\rvp$ to $\rv$ involves a coherent sum of all such paths. In this paper, the impurity is assumed more structureless
than indicated here, and corresponds to a potential $\gamma$ at the impurity site.}
\label{fig:scattering}
\end{figure}

For the case of quantum confinement, impurity scattering will be very different.\cite{BergvallPRB2013}
Consider an AGNR as illustrated in Fig.~\ref{fig:scattering} with a scattering region, here simply a localized impurity at $\rv_i$,
attached to perfect leads near $\rv$ and $\rvp$.
The leads extend far to reservoirs in the spirit of Landauer and B\"uttiker.\cite{LandauerButtiker}
The zero temperature Landauer formula reads
\begin{equation}
G(E) = G_0 \sum_{nm}T_{nm}(E),
\label{eqn:conductance}
\end{equation}
where $G_0 = 2e^2/h$ is the conductance quantum and $T_{nm}(E)$ the transmission from mode $m$ in the source electrode to $n$
in the drain. The sum is here taken over all propagating modes, i.e., all modes where the momenta $k_m(E)$ are real valued.
Note, however, that evanescent modes are included as intermediate scattering channels in the calculation of the transmission function,
i.e. evanescent modes in the scattering region are taken into account.
In general, evanescent modes extending from contacts can also be taken into account in the formulation
and the conductance will then depend on the  ratio between the distance between contact and the width of the device.\cite{TworzydloPRL2006}
Here, we consider the case when those can be neglected (system width much smaller than system length).
In a clean ribbon, free from impurities and with perfect edges, there is no mechanism for the electron to back scatter, and
the double sum reduces to a sum over the number of open modes. As function of energy, the conductance simply increases in 
integer units of the conductance quantum whenever a mode is opened. This holds both for positive energies (electron doping)
and negative energies (hole doping), where the energy is measured relative to the Dirac point.

In an unclean ribbon, the presence of impurities leads to back scattering and mode mixing. The whole matrix $T_{nm}$ becomes
relevant. In particular, interference between propagating waves and evanescent waves originating from elastic scattering at the impurities
lead to intricate interference phenomena that dominate the deviation of the conductance from the perfect step function form.

Calculations with a single localized impurity in a 1D quantum channel made from a two-dimension electron gas (2DEG)
have been done earlier \cite{Bagwell1,Bagwell2}, with the result that attractive (negative $\gamma$) impurity centers
can support quasibound states slightly below the band bottoms of transverse modes,
which leads to closing of one mode and a conductance dip of depth $G_0$.
This is a resonance phenomena between the propagating mode, say number $m$ with wave vector $k_m(E)$, and
the nearest higher evenscent mode $k_{m+1}(E)=i\kappa_{m+1}(E)$ excited by the scattering at the defect.
These reflection resonances are absent for repulsive impurities ($\gamma>0$).
At the same time, the conductance recovers exactly at the energies corresponding to the band bottoms.
This holds for both attractive and repulsive impurities.
Technically, at the band bottom $\kappa_{m+1}\rightarrow 0$ and the $T$-matrix describing multiple scattering vanishes.

For graphene, many numerical works on the problem of impurities in nanoribbons can be found
in the literature, see the reviews in Refs.~\onlinecite{RazaCollection,KatsnelsonBook,RocheBook}.
These include scaling analyses of the conductance for nanoribbons with many random impurities.
In other numerical works, conductance fluctuations due to specific types of edge disorder or scattering centers
(oxygen, nitrogen, etc.) have been presented, including different dip structures in the conductance for few scattering centers,
and loss of quantization as function of impurity density. The underlying mechanisms responsible for the resonances are
however little discussed, probably because of the heavy reliance on numerics.

The goal of this paper is to provide an analytic description for a single scattering center in a graphene ribbon
that highlight what is different in graphene with its bipartite lattice, as compared with 2DEGs,
and also highlight what is different as compared with scattering against a single scattering center in a 2D graphene sheet.
The paper can also be considered as a follow-up paper of Ref.~\onlinecite{BergvallPRB2013}, where the local density of states
and its Fourier transform (the spectral footprint) around a scattering center in an AGNR was considered in detail. Here we extend
the theory to include the signatures in electron transport (the conductance footprint).

\subsection{Scattering theory}

Let us make the above discussion a little more formal.
The probability amplitude for an electron with energy $E$ to propagate from $\rvp$ to $\rv$,
while switching from mode $m$ to $n$, is given by the electron propagator (or Green's function) $\mathbf{G}_{nm}(\rv,\rvp;E)$.
Due to the bipartite lattice of graphene, the propagators are $2\times 2$ matrices, e.g.,
\begin{equation}
\mathbf{G}_{nm}(\rv,\rvp;E) = \begin{pmatrix}
G_{nm}^{AA}(\rv,\rvp;E) & G_{nm}^{AB}(\rv,\rvp;E) \\
G_{nm}^{BA}(\rv,\rvp;E) & G_{nm}^{BB}(\rv,\rvp;E)
\end{pmatrix}.
\end{equation}
Even though the same symbol, $G$, is used also for the conductance, it should be clear to the reader which is which. The same applies
to the symbol $T$, which is both used for the transmission and (below) the $T$-matrix. To avoid cluttered
notation, the energy $E$ will sometimes be left out as an argument.
Starting with the probability amplitude $\mathbf{g}_{n}(\rv,\rvp;E)$ for an electron travelling in mode $n$ from $\rvp$ to $\rv$
in a clean ribbon, we introduce a generic localized impurity described by a $2\times 2$ matrix potential $\mathbf{\Gamma}$ and use
the Dyson equation to obtain the full electron propagator (see Appendix A),
\begin{equation}
\begin{split}
\mathbf{G}_{nm}(\rv,\rvp;E) &= \mathbf{g}_n(\rv,\rvp;E)\delta_{nm} \\
&+ \underbrace{\mathbf{g}_n(\rv,\rv_i;E)\mathbf{T}(\rv_i,\rv_i;E)\mathbf{g}_m(\rv_i,\rvp;E)}_{=\tilde{\mathbf{G}}_{nm}(\rv,\rvp;E)} \\
& = \mathbf{g}_n(\rv,\rvp;E)\delta_{nm} + \tilde{\mathbf{G}}_{nm}(\rv,\rvp;E),
\end{split}
\label{eqn:Gfull}
\end{equation}
where we call $\tilde{\mathbf{G}}_{nm}(\rv,\rvp;E)$ the scattering part. The $T$-matrix is defined as
\begin{equation}
\mathbf{T}(\rv_i,\rv_i) = \left(1 - \mathbf{\Gamma}\sum_l \mathbf{g}_l(\rv_i,\rv_i)\right)^{-1}\mathbf{\Gamma}.
\label{eqn:tmatrix}
\end{equation}
In the sum over $l$, the propagator is evaluated at the site of the impurity, $\rv_i$. The evanescent modes,
which decay exponentially away from $\rv_i$, are also included in this sum.

To calculate the conductance of a ribbon with impurities, we first note that the transmission between modes $m$ and $n$ is given by 
the scattering $s$-matrix according to $T_{nm}(E) = \left|s^{nm}_{21}(E)\right|^2$. The $s_{21}$ part of the $s$-matrix relates 
outgoing electrons in lead 2 (drain) to incoming electrons in lead 1 (source).
It can in turn be connected to the electron propagator $\mathbf{G}_{nm}(\rv,\rvp;E)$.
Such a relation, known as a Fisher-Lee relation, is derived for graphene in Appendix~\ref{app:fisherlee}, see Eq.~(\ref{eq:s-matrix}).

\section{Conductance calculations}

\begin{figure}[t]
\includegraphics[width=0.7\columnwidth]{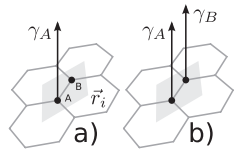}
\caption{Different impurities acting on $\rv_i$: a) single $\delta$-impurity and b) double $\delta$-impurity.}
\label{fig:impurities}
\end{figure}

For the conductance calculations,
we focus on the two impurity configurations shown in Fig.~\ref{fig:impurities}, namely single and double $\delta$-function impurities.
The single impurities are only directly connected to the A-atom in the lattice unit cell located at $\rv_i$, 
while the double site impurities affects both the A- and B-atoms.
Mathematically, the impurity potential matrix $\mathbf{\Gamma}$ is for the A-site impurity:
\begin{equation}
\mathbf{\Gamma}_A = \gamma \begin{pmatrix} 1 & 0 \\ 0 & 0 \end{pmatrix}
\end{equation}
while for the A-B-site impurity we have
\begin{equation}
\mathbf{\Gamma}_{AB} = \gamma \begin{pmatrix} 1 & 0 \\ 0 & 1 \end{pmatrix}.
\end{equation}
%
%
%

In our earlier work,\cite{BergvallPRB2013} the free propagator $\mathbf{g}_{n}(\rv,\rvp;E)$ and the propagator
including a localized A-site impurity $\mathbf{G}_{nm}(\rv,\rvp;E)$ were given explicitly.
For the results below, we have also computed $\mathbf{G}_{nm}(\rv,\rvp;E)$
for the A-B-site impurity, but it is a quite long expression and not worth reprinting here. The calculation of it is straight forward starting
with the formula in Appendix A and the procedure outlined in Ref.~\onlinecite{BergvallPRB2013}.

\subsection{A-site impurity}

\begin{figure}[t]
\includegraphics[width=\columnwidth]{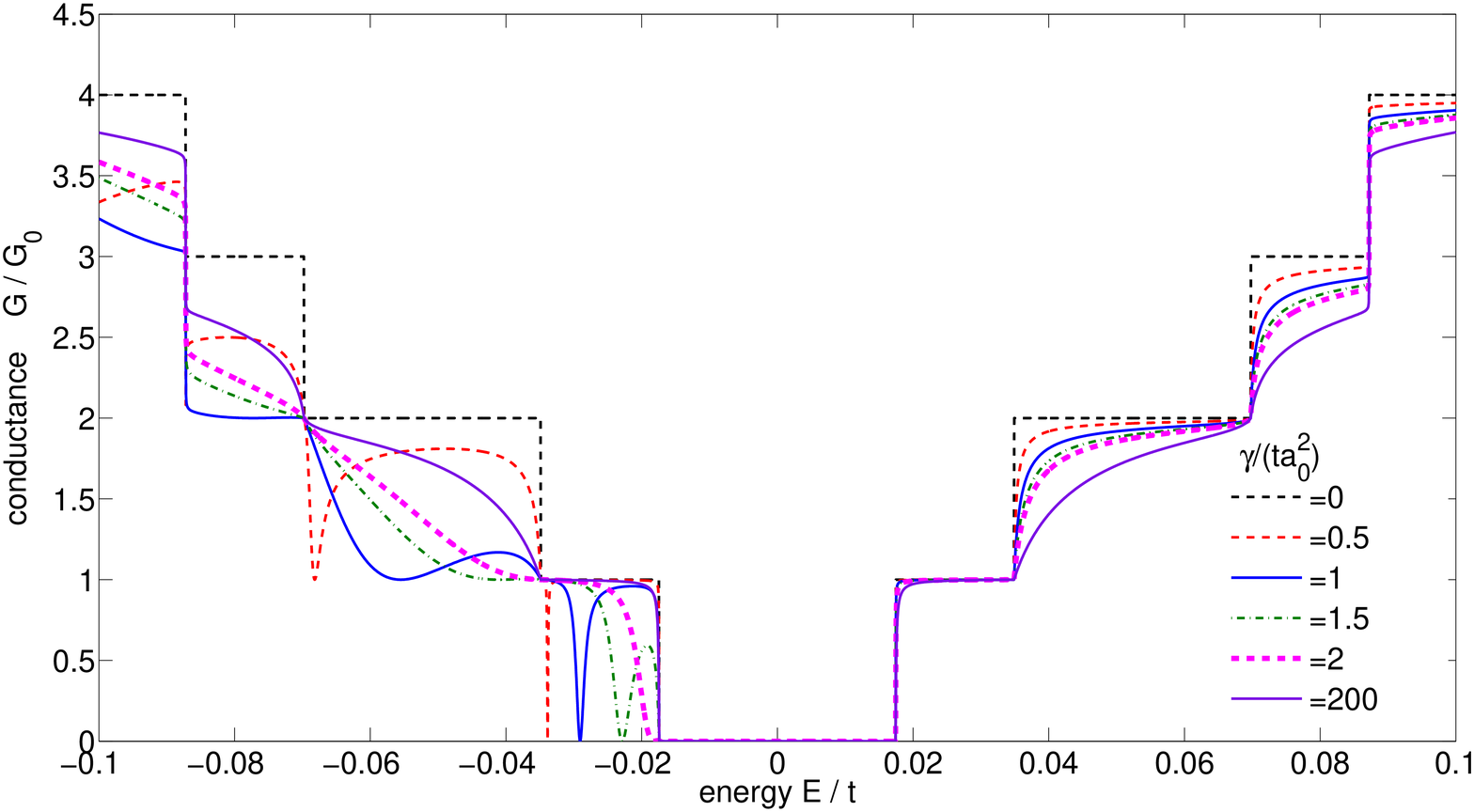}
\caption{A-site impurity: The zero temperature conductance in unit of the conductance quantum for a AGNR $102$ atoms wide for
varying repulsive A-site impurity strengths $\gamma$ for an impurity located at $x_i=W/1.73$ from the ribbon edge.}
\label{fig:impurityA}
\end{figure}

For the single $\delta$-function impurity, as is shown in Fig.~\ref{fig:impurities}(a), we display the conductance in Fig.~\ref{fig:impurityA}.
Two main features are apparent in the figure.
First, there are characteristic dips in the conductance for negative energies, where exactly one conductance quantum is lost.
Secondly, the plateaux conductances exactly at the energies where new modes open appear to remain largely intact,
especially on the positive energy side.
The origin of all these features can be studied analytically.

We can write the $T$-matrix as\cite{BergvallPRB2013}
\begin{equation}
T(\rv_i,\rv_i;E) = \frac{\gamma}{1 + \gamma\sigma_e(E)  + i\gamma\sigma_p(E)} \begin{pmatrix} 1 & 0 \\ 0 & 0 \end{pmatrix},
\label{eq:T_A}
\end{equation}
where the subscript $e$ and $p$ denotes evanescent and propagating modes, respectively, and
\begin{equation}
\begin{split}
\sigma_p(E) &= \frac{|E|}{(\hbar v_f)^2}\sum_{l}\frac{\chi_l(x_i)}{\kappa_l(E)}\\
\sigma_e(E) &= \frac{E}{(\hbar v_f)^2}\sum_{l}\frac{\chi_l(x_i)}{\kappa_l(E)}.
\end{split}
\label{eq:sigma}
\end{equation}
These functions depend on energy $E$, the transverse eigenmodes at the location of the impurity
$\chi_n(x_i)=\sqrt{2/W}\sin\left[(n\pi/W)x_i\right]$ ($W$ is the ribbon width), and the magnitudes of
the longitudinal momenta, defined as
$\kappa_n(E)=\sqrt{|E^2/(\hbar v_f)^2 - q_n^2|}$,
where $q_n=n\pi/W-4\pi/3a$ is the (quantized) transverse momentum in mode $n$.
That the evanescent modes actually have imaginary longitudinal wavevectors $k_n^e(E) = i\mbox{sgn}(E)\kappa_n(E)$
has been taken into account above.

From Eq.~(\ref{eq:T_A}) it is clear that at the bottom of a subband $E_{n+1}(0)$,
where the wave vector of the evanescent mode $n+1$ goes to zero,
$\sigma_e(E)$ diverges and the T-matrix vanishes, since $\sigma_p$ remains finite. Therefore, all conductance plateaux
values $G=nG_0$ at energies corresponding to the bottom of subbands $n+1$ remains intact for this type of impurity.
Also, if the equation $1+\gamma\sigma_e(E)=0$ can be fulfilled, there is a posibility of a transmission resonance.
Both results are analagous to the situation in the 1D quantum channel made from
a 2DEG,\cite{Bagwell1,Bagwell2} where it was shown that the conductance steps are always present at the bottom
of subbands and dips can occur when the impurity potential is attractive. In the case of graphene, the repulsive impurity 
corresponds to an attractive one on the hole-side of the spectrum (negative energies) and dips are found on that side.
We note that if we change sign of $\gamma$, the whole picture in Fig.~\ref{fig:impurityA} is flipped $E\rightarrow -E$.

To quantify the resonance behavior further, we keep one propagating mode denoted $n$ and one evanescent mode
denoted $n+1$ and compute the position of the dip relative to the bottom of the evanescent subband $E_{n+1}(0)$.
We obtain
\begin{equation}
\Delta E = E_{n+1}(0) \left|1 - \frac{1}{|\gamma|}\frac{\hbar v_f}{\sqrt{\chi_e^4(x_i) + (\hbar v_f/\gamma)^2}}\right|. 
\end{equation}
It can be shown that at this energy, the scattering part of the propagator
$\tilde{\mathbf{G}}_{nn}(\rv,\rvp;E_{dip}) = -\mathbf{g}_n(\rv,\rvp;E_{dip})$ and the full propagator $\mathbf{G}_{nn}(\rv,\rvp;E_{dip})$
between source and drain vanishes and the contribution to the conductance from this channel is lost.
For small $\gamma$ the resonance is close to the conductance step, while for intermediate $\gamma$ it can 
be located anywhere along the plateau. For very large $\gamma$, the dip merges with the step, but the effect discussed
above for vanishing $\kappa^e_{n+1}$ wins and the step structure remains intact. The conductance becomes
$E\rightarrow -E$ symmetric as $\gamma\rightarrow\infty$ and there are no dips.
The above formula is exact for the lowest open mode but should be corrected for higher modes. Depending on
the value of $\gamma$ some resonances are not perfect back-reflection resonances on higher plateaux,
which is also clear from Fig.~\ref{fig:impurityA}.

\begin{figure}
\includegraphics[width=\columnwidth]{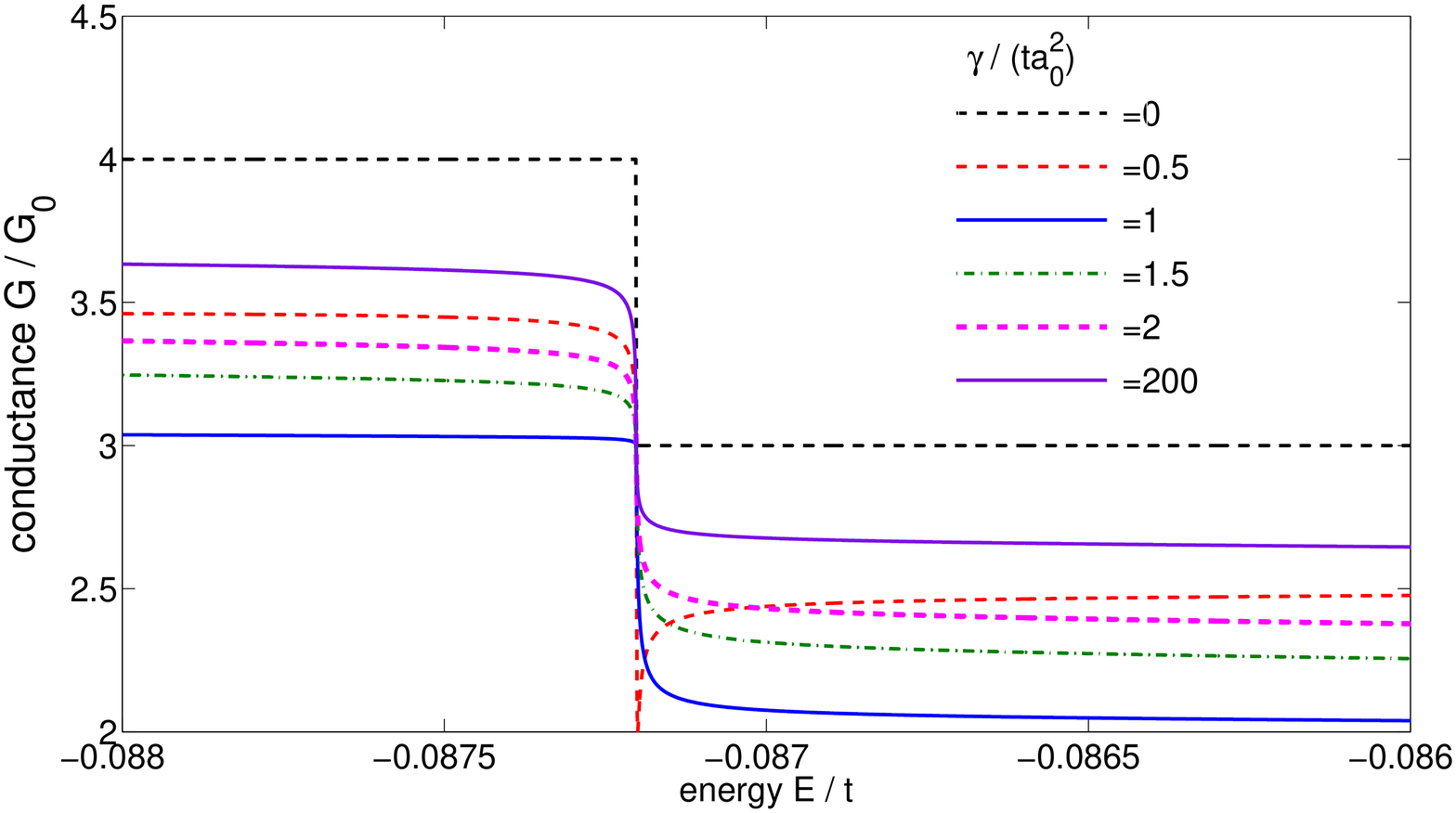}
\caption{Zoom in of the conductance in Fig.~\ref{fig:impurityA} at one of the conductance steps.}
\label{fig:impurityA_zoom1}
\end{figure}

\begin{figure}
\includegraphics[width=\columnwidth]{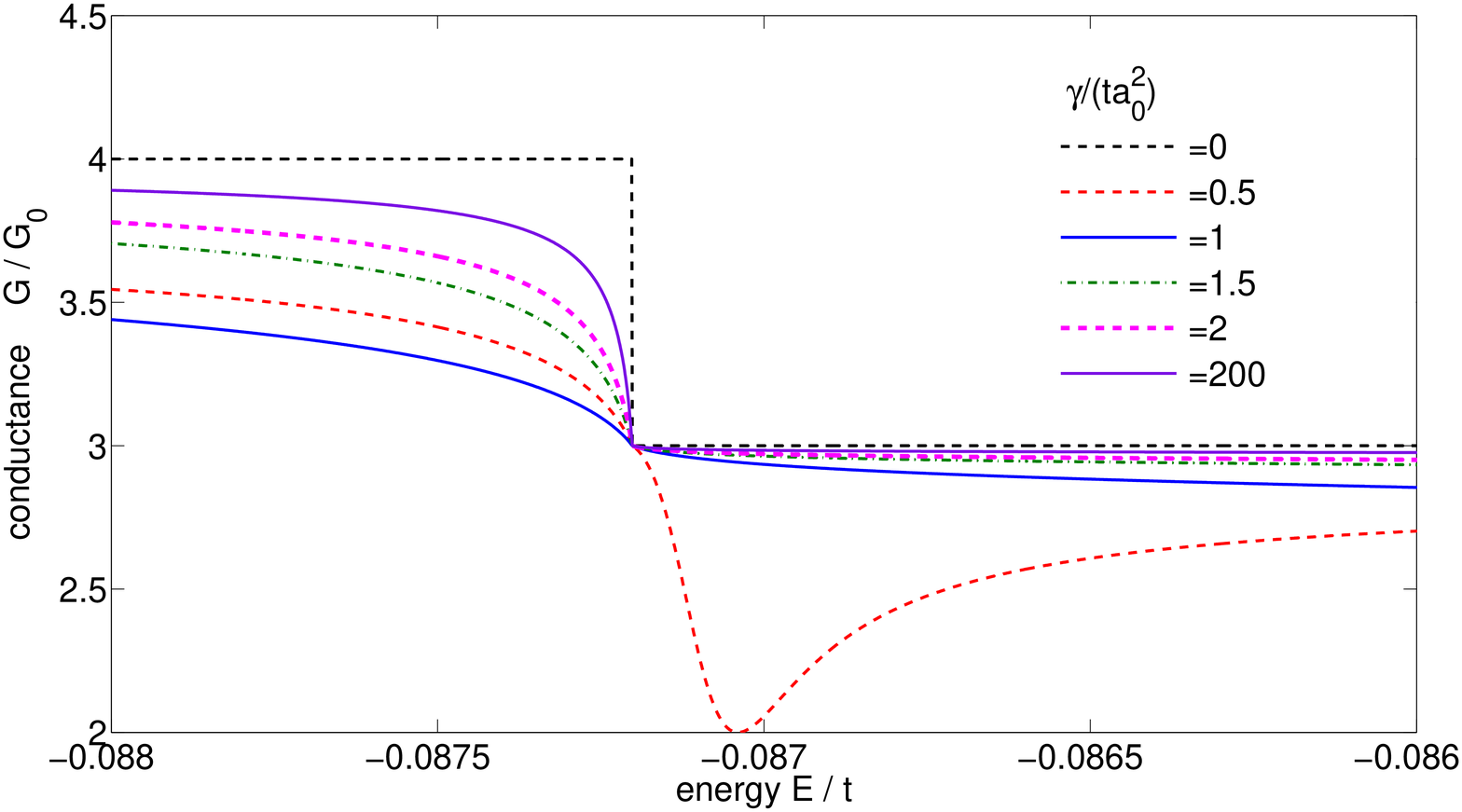}
\caption{The conductance for energies as in Fig.~\ref{fig:impurityA_zoom1} at one of the conductance steps
but for the impurity located at $x_i=0.1W$.}
\label{fig:impurityA_zoom2}
\end{figure}

If we inspect Fig.~\ref{fig:impurityA} more carefully, we start to suspect that some steps are lost, for instance the one
on the $G=3G_0$ where the fourth mode opens. But if we zoom in and look carefully, as in Fig.~\ref{fig:impurityA_zoom1},
we see that the conductance $3G_0$ at the point when the next higher lying subband (the fourth) is opened remains intact, although
the recovery is very steep. Actually, this is accidental, in that the impurity position was chosen as $x_i=W/1.73$. If we choose another
impurity position, the transparency of the impurity and the conductance recovery is more clearly visible, see Fig.~\ref{fig:impurityA_zoom2}

\subsection{A-B-site impurity}

\begin{figure}
\includegraphics[width=\columnwidth]{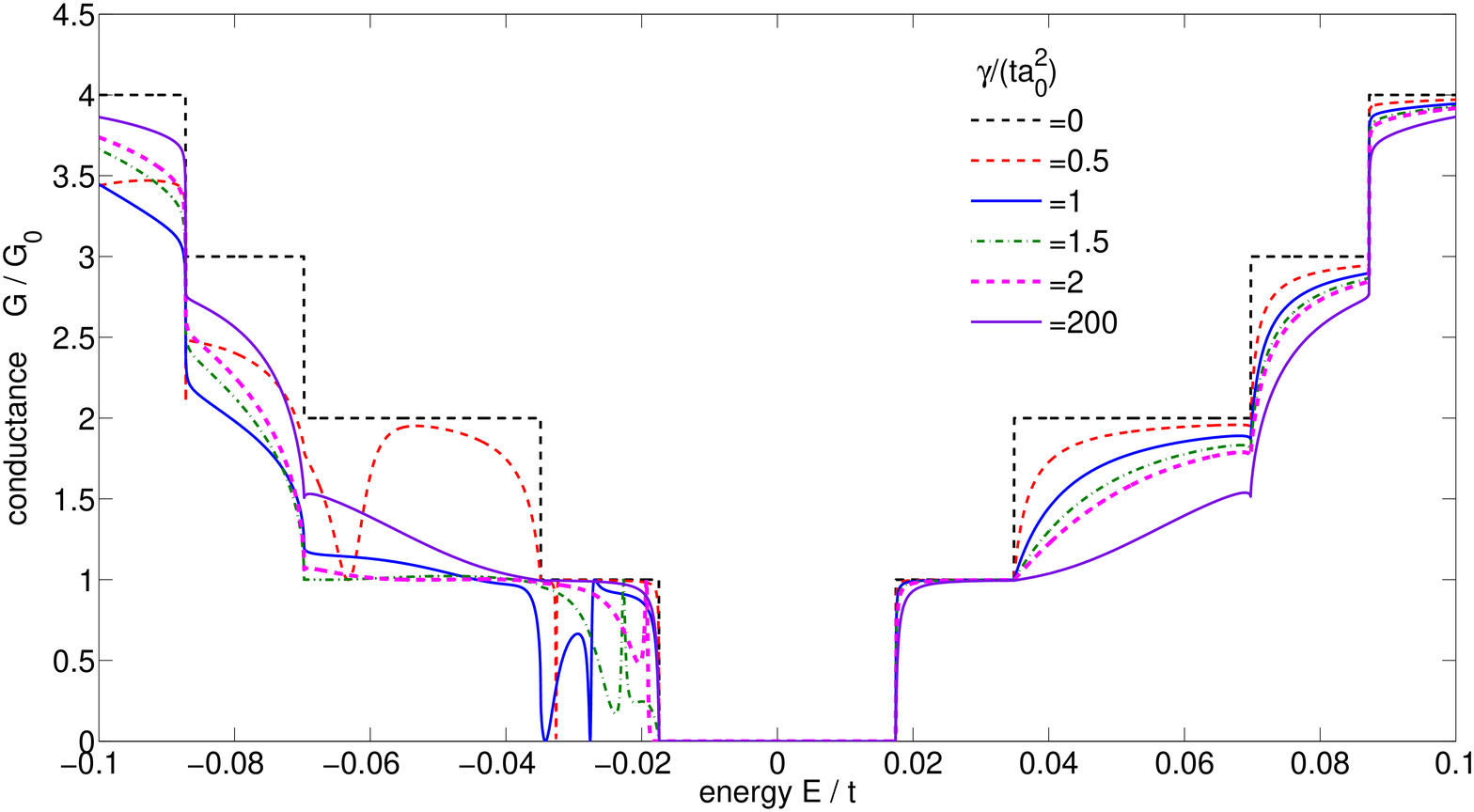}
\caption{A-B-site impurity: The zero temperature conductance in unit of the conductance quantum for a AGNR $102$ atoms wide for
varying repulsive A-B-site impurity strengths $\gamma$ for an impurity located at $x_i=W/1.73$ from the ribbon edge.}
\label{fig:impurityAB}
\end{figure}

\begin{figure}
\includegraphics[width=\columnwidth]{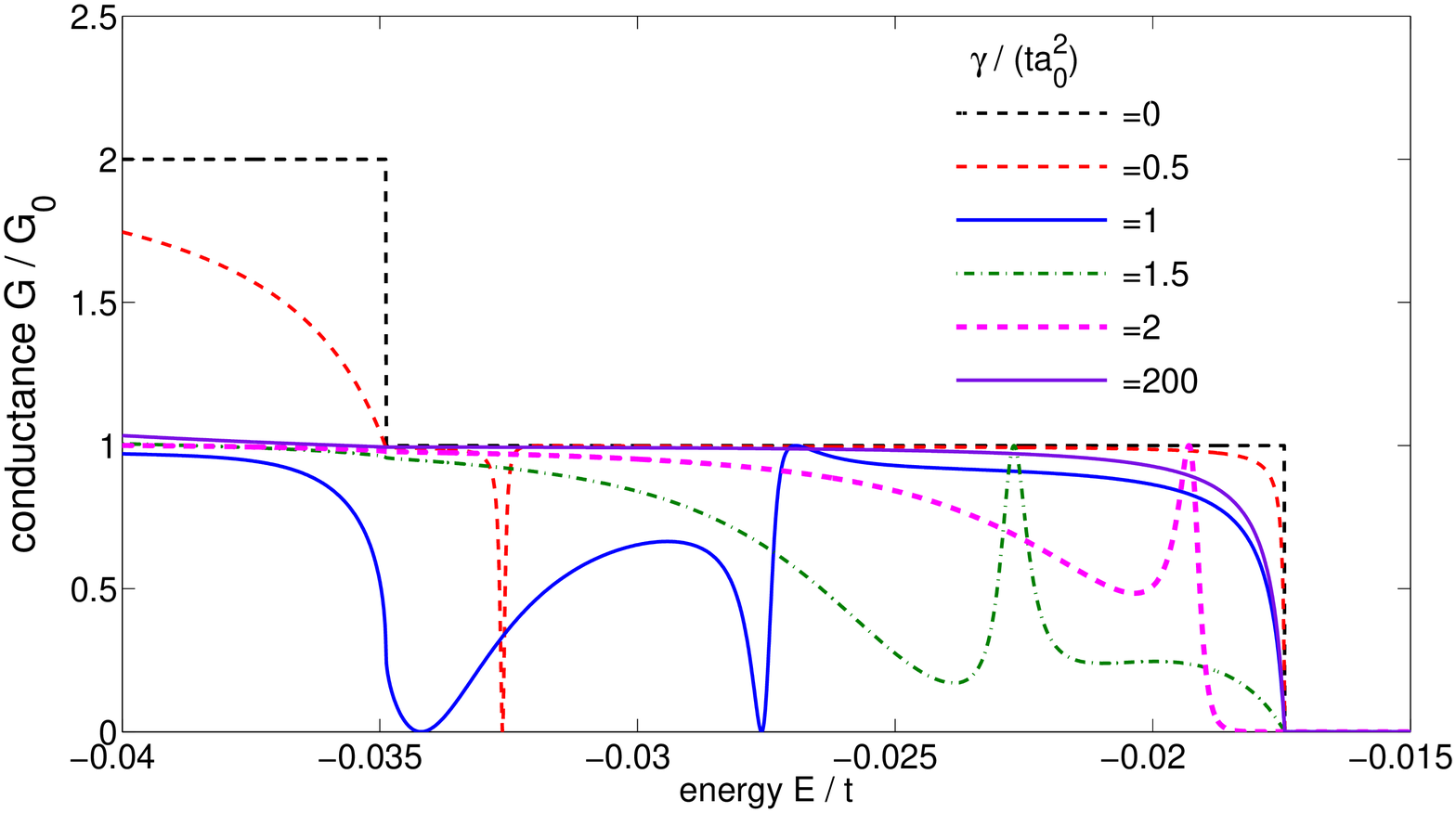}
\caption{Zoom in of the conductance in Fig.~\ref{fig:impurityAB} around the first conductance step.}
\label{fig:impurityAB_zoom}
\end{figure}

Next we consider the case of a scalar impurity in the $A-B$ sublattice space (diagonal potential), a situation corresponding
to an impurity potential on two neighboring sites, as illustrated in Fig.~\ref{fig:impurities}(b). We display the conductance
for varying impurity strengths $\gamma$ in Fig.~\ref{fig:impurityAB}. As is clear from the figure, the situation is more complicated
as compared with the A-site impurity. We zoom in on the structures around the first step in Fig.~\ref{fig:impurityAB_zoom} and
concentrate the discussion around this figure.

This case is unique to graphene and has no direct correspondence for a 1D channel made from a 2DEG.
The off-diagonal components of the A-B sublattice matrix propagator $\mathbf{g}_{n}(\rv_i,\rv_i$;E)
enters directly in the $T$-matrix equation. These components contain factors encoding
the Dirac electron chirality, see Eqs.~(B5)-(B10) in Ref.~\onlinecite{BergvallPRB2013}.
The full analytic expression for the $T$-matrix is rather lengthy and not so informative. Therefore,
we write it in a short-hand notation to display its structure,
\begin{equation}
T(E) = \frac{\gamma}{(1-\gamma\tilde{g}_{AA})^2 - (\gamma\tilde{g}_{AB})^2} \begin{pmatrix} 1-\gamma\tilde{g}_{AA} & \gamma\tilde{g}_{AB} \\ \gamma\tilde{g}_{AB} & 1-\gamma\tilde{g}_{AA}\end{pmatrix}.
\end{equation}
For instance, the function $\tilde{g}_{AA}$ reduces to $\sigma_e+i\sigma_p$ for the A-site impurity in Eq.~(\ref{eq:T_A}) above.
The function $\tilde{g}_{AB}$ corresponds to interference between the A-site and B-site scattering events through the opportunity
of changing sublattice. The $AA$-element of this $T$-matrix can be rewritten as
\begin{equation}
\begin{split}
T_{AA}(E) &= \frac{\gamma(1-\gamma\tilde{g}_{AA})}{(1-\gamma\tilde{g}_{AA})^2 - (\gamma\tilde{g}_{AB})^2} \\
&= \frac{\gamma}{2}\left[\frac{1}{1-\gamma(\tilde{g}_{AA} - \tilde{g}_{AB})} + \frac{1}{1 - \gamma(\tilde{g}_{AA} + \tilde{g}_{AB})}\right].
\end{split}
\end{equation}
It can be shown that
\begin{equation}
\tilde{g}_{AA} \pm \tilde{g}_{AB} \propto \sum_{n} \frac{1}{\kappa_n}\left(|q_n| \pm  q_n\right),
\end{equation}
and the first fraction in $T_{AA}(E)$ will stay finite when a $\kappa_n \rightarrow 0$, while the second one will approach zero.
Thus, the $T$-matrix stays finite and we will not regain a perfect channel as $\kappa_n\rightarrow 0$ as we did in the A-site impurity case.
This is due to the extra freedom of changing sublattice during multiple scattering events at the A and the B site impurities.
In addition to this, there are more degrees of freedom to aquire resonances in the $T$-matrix which is apparent from the rich
resonance structure in Fig.~\ref{fig:impurityAB_zoom}. In this case, both transmission dips (including a complete loss of a channel)
and transmission resonance in the middle of the plateau (green dash-dotted line in the figure) can occur depending on the value of $\gamma$.

\section{Summary}

In summary, we have presented analytical results for electron transport in graphene nanoribbons with perfect armchair edges. We 
have shown how the interplay between evanescent waves around localized scattering centers and propagating waves connecting
source and drain generates
back-scattering resonances that result in dips in the transmission function.
At the same time back-scattering is completely suppressed for the A-site impurity at energies corresponding to
the bottom of subbands, where the conductance completely recovers.
This is analagous to the case of 1D channels made from 2DEGs. However, for an A-B site impurity, 
multiple scattering at the double impurity includes intermediate A to B sublattice scattering that interfere with
the transmission resonance and it is lost.
We have derived a Fisher-Lee type expression relating the transmission to the electron propagator that can be
used for further studies of more complicated scattering regions.

\section{Acknowledgements}

This work has been supported by the Swedish Foundation for Strategic Research (SSF),
Knut and Alice Wallenberg foundation (KAW),
and the EU through the FP7 project ConceptGraphene.

\appendix

%
%

\section{T-matrix equation}\label{app:tmatrix}

By looking at Fig.~\ref{fig:scattering}, we see that the propagator from $\rvp$ to $\rv$ may be written as
\begin{equation}
\begin{split}
G_{nm}^{\rv,\rvp} &= g_{n}^{\rv,\rvp}\delta_{nm} + g_{n}^{\rv,\rv_i}\Gamma_{nm}^{\rv_i,\rv_i}g_{m}^{\rv_i,\rvp} \\
&+ \sum_{l}g_{n}^{\rv,\rv_i}\Gamma_{nl}^{\rv_i,\rv_i}g_{l}^{\rv_i,\rv_i}\Gamma_{lm}^{\rv_i,\rv_i}g_{m}^{\rv_i,\rvp} \\
&+ \sum_{ll^\prime}g_{n}^{\rv,\rv_i}\Gamma_{nl}^{\rv_i,\rv_i}g_{l}^{\rv_i,\rv_i}\Gamma_{ll^\prime}^{\rv_i,\rv_i}g_{l^\prime}^{\rv_i,\rv_i}\Gamma_{l^\prime,m}^{\rv_i,\rv_i}g_{m}^{\rv_i,\rvp} + \ldots\\
&= g_n^{\rv,\rvp}\delta_{nm} + \sum_{l}g_n^{\rv,\rv_i}\Gamma_{nl}^{\rv_i,\rv_i}G_{lm}^{\rv_i,\rvp},
\end{split}
\end{equation}
where we have introduced the shorthand notation $A_{nm}^{\rv,\rvp} = A_{nm}(\rv,\rvp;E)$,
and where $\Gamma_{nm}^{\rv_i,\rv_i}$ corresponds to the
process of the electron hopping from the impurity site $\rv_i$, interacting with the impurity and then hopping back.
In the process, the electron is scattered from mode $m$ to $n$.
If we assume that the potential does not depend on mode  index or position,
we have that $\Gamma_{nm}^{\rv_i,\rv_i} = \Gamma$. Still, the A-B sub lattice degree of freedom remains
and $\Gamma$ is a $2\times 2$ matrix.

By observation, the expression for the propagator may be rewritten as
\begin{equation}
G_{nm}^{\rv,\rvp} = g_n^{\rv,\rvp}\delta_{nm} + g_n^{\rv,\rv_i}T_{nm}^{\rv_i,\rv_i}g_m^{\rv_i,\rvp},
\end{equation}
where the $T$-matrix is
\begin{equation}
\begin{split}
T_{nm}^{\rv_i,\rv_i} &= \Gamma + \sum_l \Gamma g_l^{\rv_i,\rv_i}\Gamma + \sum_{ll^\prime}\Gamma g_l^{\rv_i,\rv_i}\Gamma g_{l^\prime}^{\rv_i,\rv_i}\Gamma + \ldots \\
&= \Gamma + \Gamma\sum_l g_l^{\rv_i,\rv_i} T_{nm}^{\rv_i,\rv_i},
\end{split}
\end{equation}
or,
\begin{equation}
T_{nm}^{\rv_i,\rv_i} = \left(1 - \Gamma\sum_l g_l^{\rv_i,\rv_i}\right)^{-1} \Gamma.
\end{equation}
Since there is no mode-dependence in the right-hand side, we can drop the mode-indices from the $T$-matrix, and we arrive at the 
final expression,
\begin{equation}
T^{\rv_i,\rv_i} = \left(1 - \Gamma \sum_l g_l^{\rv_i,\rv_i}\right)^{-1}\Gamma.
\end{equation}

%
%

%
%

\section{Derivation of a Fisher-Lee transport formula for graphene}\label{app:fisherlee}

\begin{figure}[t]
\includegraphics[width=\columnwidth]{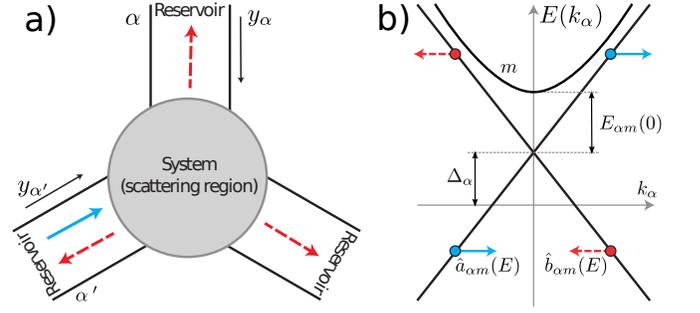}
\caption{(a) A sketch of the system, divided into scattering region and leads.
(b) Example of the dispersion relations $E_{\alpha m}(k_{\alpha})$ for a metallic AGNR,
including the definition of $\Delta_{\alpha}$ as a shift of the Dirac point away from zero by a backgate voltage capacatively
coupled to the graphene lead. The band bottoms are denoted $E_{\alpha m}(0)$. Also indicated are operators annihilating
{\it incoming} ($\hat a$, blue arrows) and {\it outgoing} ($\hat b$, red arrows) waves in the lead, where incoming and outgoing
is defined with respect to the scattering region with the coordinate systems defined in (a).}
\label{fig:system}
\end{figure}

In this appendix, we derive an expression relating the transmission function between graphene leads to the 
electron propagator (Green's function) of the graphene scattering region connecting the leads together,
as displayed in Fig.~\ref{fig:system}(a).
When we need to be specific, we use the AGNR waveguide lead eigenfunctions which are known analytically,
but the formalism can be generalized to other leads as long as one at least numerically can compute the lead eigenfunctions.
Following B\"uttiker,\cite{ButtikerPRB1992} we denote by $T_{\alpha \alpha^\prime, m m^\prime}(E)$ the probability of an electron wave
incoming in lead $\alpha^\prime$  in transverse mode $m^\prime$, to exit through lead $\alpha$ in transverse mode $m$.
We study only elastic scattering processes, and $E$ is thus the energy of both the incoming and outgoing waves.

Using a local coordinate system in each lead, where $x_\alpha$ is the transverse coordinate and $y_\alpha$ the longitudinal one, 
increasing in the direction towards the scattering region, we write down the wavefunction of an electron in lead $\alpha$ and 
transverse mode $m$ as
\begin{equation}
\vec{\psi}_{\alpha m \nu}(E) = \chi_{\alpha m}(x_\alpha) \vec{\phi}_{\alpha m \nu}(E) e^{i\nu\sgn{E-\Delta_\alpha} k_{\alpha m}(E) y_\alpha},
\end{equation}
where the transverse wave functions for AGNRs are given by
\begin{equation}
\chi_{\alpha m}(x_\alpha) = \sqrt{\frac{2}{W_\alpha}}\sin\left(\frac{n\pi}{W_\alpha}x_{\alpha}\right),
\end{equation}
where $W_{\alpha}$ is the width of lead $\alpha$. The pseudo spinor is given by
\begin{equation}
\vec{\phi}_{\alpha m \nu}(E) = \begin{pmatrix} \frac{\nu k_{\alpha m}(E) + i\sgn{E-\Delta_\alpha}q_{\alpha m}}{\sqrt{k_{\alpha m}^2(E) + q_{\alpha m}^2}} \\ i \end{pmatrix}.
\end{equation}
Above, $\nu = \pm$ denotes an incoming or outgoing wave, as indicated in Fig.~\ref{fig:system}(b).
The transverse momenta are defined as $q_{\alpha m} = m\pi/W_\alpha - 4\pi/3$,
while the longitudinal momenta are given by
\begin{equation}
k_{\alpha m}(E) = \frac{1}{\hbar v_f}\sqrt{(E-\Delta_\alpha)^2 - E_{\alpha m}^2(0)},
\end{equation}
where $v_f$ the Fermi velocity of pristine two-dimensional graphene,
$E_{\alpha m}(0)$ is the band bottom of mode $m$,
and $\Delta_\alpha$ is a possible shift of the Dirac point away from zero in lead $\alpha$
by application of a back gate voltage, as also indicated in Fig.~\ref{fig:system}(b).

We introduce the operators $\hat{a}_{\alpha m}(E)$ and $\hat{b}_{\alpha m}(E)$, that annihilates an incoming and outgoing states,
respectively, in lead $\alpha$ and mode $m$ (see Fig.~\ref{fig:system}b). The general scattering state in lead $\alpha$ can then
be written as
\begin{widetext}
\begin{equation}
\hat{\Psi}_{\alpha}(\rv,t) = \sum_m \int_{-\infty}^\infty \frac{dE}{2\pi} \sqrt{\frac{E-\Delta_\alpha}{(\hbar v_f)^2k_{\alpha m}(E)}}
e^{-iEt/\hbar} \left[\hat{a}_{\alpha m}(E)\vec{\psi}_{\alpha m +}(E) + \hat{b}_{\alpha m}(E)\vec{\psi}_{\alpha m -}(E)\right] \Theta_{\alpha m}(E),
\label{eq:psi_alpha}
\end{equation}
\end{widetext}
where $\Theta_{\alpha m}(E) = \theta(E-[\Delta_\alpha + E_{\alpha m}(0)]) + \theta([\Delta_\alpha -E_{\alpha m}(0)] - E)$
is added to make sure that we only incorporate propagating transverse eigenmodes in the leads. The first and second
step functions correspond to the electron bands (above the Dirac point) and hole bands (below the Dirac point), respectively

By definition, the retarded Green's function, propagating an electron from lead $\alpha^\prime$ to lead $\alpha$, is
\begin{equation}
\mathbf{G}_{\alpha \alpha^\prime}(\rv, \rvp;t,t^\prime) = 
-i\Theta(t-t^\prime)\left\langle \left\{\hat{\Psi}_{\alpha}(\rv,t), \hat{\Psi}_{\alpha^\prime}^\dagger(\rvp,t^\prime)  \right\}\right\rangle.
\end{equation}
Using Eq.~(\ref{eq:psi_alpha}), we find that
\begin{widetext}
\begin{equation}
\begin{split}
\mathbf{G}_{\alpha \alpha^\prime}(\rv, \rvp;t,t^\prime) &= -i\Theta(t-t^\prime) \sum_{m m^\prime} 
\int_{-\infty}^\infty dE \int_{-\infty}^\infty dE^\prime e^{-i(Et - E^\prime t^\prime)/\hbar} \Theta_{\alpha m}(E)\Theta_{\alpha^\prime m^\prime}(E^\prime)\\
&\times  
\sqrt{\frac{E-\Delta_\alpha}{(\hbar v_f)^2 k_{\alpha m}(E)}} \sqrt{\frac{E^\prime-\Delta_{\alpha^\prime}}{(\hbar v_f)^2 k_{\alpha^\prime m^\prime}(E^\prime)}} \\
&\times \left[\;\;
				\vec{\psi}_{\alpha m +}(E)\vec{\psi}^\dagger_{\alpha^\prime m^\prime +}(E^\prime) \left\langle \left\{ \hat{a}_{\alpha m}(E), \hat{a}^\dagger_{\alpha^\prime m^\prime}(E^\prime)\right\} \right\rangle       \right. \\
&\quad + 		\vec{\psi}_{\alpha m +}(E)\vec{\psi}^\dagger_{\alpha^\prime m^\prime -}(E^\prime) \left\langle \left\{ \hat{a}_{\alpha m}(E), \hat{b}^\dagger_{\alpha^\prime m^\prime}(E^\prime)\right\} \right\rangle \\
&\quad + 		\vec{\psi}_{\alpha m -}(E)\vec{\psi}^\dagger_{\alpha^\prime m^\prime +}(E^\prime) \left\langle \left\{ \hat{b}_{\alpha m}(E), \hat{a}^\dagger_{\alpha^\prime m^\prime}(E^\prime)\right\} \right\rangle \\
&\quad \left. + \vec{\psi}_{\alpha m -}(E)\vec{\psi}^\dagger_{\alpha^\prime m^\prime -}(E^\prime) \left\langle \left\{ \hat{b}_{\alpha m}(E), \hat{b}^\dagger_{\alpha^\prime m^\prime}(E^\prime)\right\} \right\rangle \right].
\end{split}
\end{equation}
\end{widetext}
Next we need the commutation relations for the operators. First, we have
\begin{equation}
\begin{split}
\left\{\hat{a}_{\alpha m}(E), \hat{a}^\dagger_{\alpha^\prime m^\prime}(E^\prime)\right\} &= \delta_{\alpha \alpha^\prime}^{m m^\prime}\delta(E-E^\prime), \\
\left\{\hat{b}_{\alpha m}(E), \hat{b}^\dagger_{\alpha^\prime m^\prime}(E^\prime)\right\} &= \delta_{\alpha \alpha^\prime}^{m m^\prime}\delta(E-E^\prime).
\end{split}
\end{equation}
The outgoing waves are related to the incoming waves through the scattering matrix as
\begin{equation}
\hat{b}_{\alpha m}(E) = \sum_{\beta n} s_{\alpha \beta, m n}(E) \hat{a}_{\beta n}(E),
\end{equation}
which leads to
\begin{equation}
\left\{ \hat{a}_{\alpha m}(E), \hat{b}^\dagger_{\alpha^\prime m^\prime}(E^\prime) \right\} = s^\dagger_{\alpha^\prime \alpha, m^\prime m}(E)\delta(E-E^\prime)
\end{equation}
and
\begin{equation}
\left\{ \hat{b}_{\alpha m}(E), \hat{a}^\dagger_{\alpha^\prime m^\prime}(E^\prime)\right\} = s_{\alpha \alpha^\prime, m m^\prime}(E)\delta(E-E^\prime) .
\end{equation}
By using these commutation relations, the expression for the propagator turns into
\begin{widetext}
\begin{equation}
\begin{split}
\mathbf{G}_{\alpha \alpha^\prime}(\rv, \rvp;t,t^\prime) &= -i\Theta(t-t^\prime) \sum_{m m^\prime} 
\int_{-\infty}^\infty dE e^{-iE(t - t^\prime)/\hbar} \Theta_{\alpha m}(E)\Theta_{\alpha^\prime m^\prime}(E)\\
&\times  
\sqrt{\frac{E-\Delta_\alpha}{(\hbar v_f)^2 k_{\alpha m}(E)}} \sqrt{\frac{E-\Delta_{\alpha^\prime}}{(\hbar v_f)^2 k_{\alpha^\prime m^\prime}(E)}} \\
&\times \left[\;\;
				\vec{\psi}_{\alpha m +}(E)\vec{\psi}^\dagger_{\alpha m +}(E) \delta_{\alpha \alpha^\prime}^{m m^\prime}   + \vec{\psi}_{\alpha m +}(E)\vec{\psi}^\dagger_{\alpha^\prime m^\prime -}(E) s^\dagger_{\alpha^\prime \alpha, m^\prime m}(E)    \right. \\
&\quad \left. + \vec{\psi}_{\alpha m -}(E)\vec{\psi}^\dagger_{\alpha m -}(E) \delta_{\alpha \alpha^\prime}^{m m^\prime}   + \vec{\psi}_{\alpha m -}(E)\vec{\psi}^\dagger_{\alpha^\prime m^\prime +}(E) s_{\alpha \alpha^\prime, m m^\prime}(E)\right].
\end{split}
\label{eqn:Gbig}
\end{equation}
\end{widetext}
We multiply \eq{eqn:Gbig} by the transverse wavefunctions $\chi_n(x)$ and $\chi_{n^\prime}(x^\prime)$
from the left and right, respectively, and integrate the transverse coordinates $x$ and $x^\prime$ over the ribbon widths.
We then kill the double sum over transverse modes leaving us with
a propagator $\mathbf{G}_{\alpha \alpha^\prime}^{n n^\prime}(y,y^\prime;t,t^\prime)$,
for scattering from mode $n^\prime$ to $n$.

For this stationary problem, we only have a dependence on time-difference. It is then natural to introduce $\tau = t-t^\prime$
and integrate over $\tau$ to obtain the Fourier-transform of the propagator,
$\mathbf{G}_{\alpha \alpha^\prime}^{n n\prime}(y,y^\prime;\hbar\omega)$.
Moving out factors not depending on $\tau$ in front of the integral, we get (leaving out the prefactor for now) the standard integral,
\begin{equation}
-i \int_0^\infty d\tau e^{i(\hbar\omega -E + i\eta)\tau/\hbar} = \frac{\hbar}{\hbar\omega - E  + i\eta},
\label{eqn:Gtemp}
\end{equation}
where $\eta$ is a small positive number.

\begin{figure}[t]
\includegraphics[width=\columnwidth]{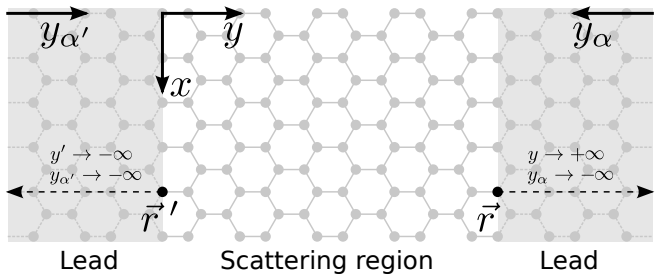}
\caption{Definition of the various longitudinal AGNR coordinates needed in derivation of the
Fisher-Lee formula, see text in connection with Eq.~(\ref{eq:G_coord}).}
\label{fig:coordinates}
\end{figure}

At this point it is time to perform the integral over energy $E$.
Since the integrand has a single pole at $E = \hbar\omega + i\eta$ in the upper complex half-plane,
it is natural to move into the complex plane ($E \rightarrow z$),
and integrate slightly above or below the $E$-axis and close the contour either in the upper or lower half of the complex plane,
depending on the convergence criterium on the closing arc at large distance $|z|=R\rightarrow\infty$,
which is controlled by the signs of $k_{\alpha}(z)$ in the different
terms. It turns out that only the last two terms in Eq.~(\ref{eqn:Gbig}) will allow us to close the contour in the upper plane where the pole exists
and only those two terms contribute. We obtain
\begin{widetext}
\begin{equation}
\begin{split}
\mathbf{G}_{\alpha  \alpha^\prime}^{n n^\prime}(y,y^\prime;\hbar\omega) 
&= -ih \sqrt{\frac{\hbar \omega - \Delta_\alpha}{(hv_f)^2k_{\alpha n}(\hbar\omega)}} \sqrt{\frac{\hbar \omega - \Delta_{\alpha}}{(hv_f)^2k_{\alpha^\prime n^\prime}(\hbar\omega)}} \\
&\times \left[  \vec{\phi}_{\alpha n -}(\hbar\omega) \vec{\phi}^\dagger_{\alpha n -}(\hbar\omega) e^{i\sgn{\hbar\omega-\Delta_{\alpha}} k_{\alpha n}(\hbar\omega)|y_\alpha - y^\prime_{\alpha}|} \delta_{\alpha\alpha^\prime}^{n n^\prime}   \right. \\
&~\quad + \left.  \vec{\phi}_{\alpha n -}(\hbar\omega) \vec{\phi}^\dagger_{\alpha^\prime n^\prime +}(\hbar\omega) e^{i\sgn{\hbar\omega-\Delta_{\alpha}} k_{\alpha n}(\hbar\omega)|y_\alpha|}e^{i \sgn{\hbar\omega-\Delta_{\alpha^\prime}} k_{\alpha^\prime n^\prime}(\hbar\omega)|y^\prime_{\alpha^\prime}|} s_{\alpha\alpha^\prime,n n^\prime}(\hbar\omega)  \right].
\end{split}
\label{eq:G_coord}
\end{equation}
We now have an expression relating the propagator to the $s$-matrix, but we want the inverse relation.
To obtain the $s$-matrix we need to project with the ribbon pseudo-spinors, and divide down the prefactors.
In that last step one can get rid of the spurious coordinate dependence in Eq.~(\ref{eq:G_coord}) by a subtle trick.\cite{FisherLee}
One needs to observe that the coordinate dependence of the propagator
$\mathbf{G}_{\alpha  \alpha^\prime}^{n n^\prime}(y,y^\prime;\omega)$ also is in the form of exponential functions, that exactly
cancel the ones we divide down.\footnote{For our case of an AGNR with a defect being our system,
this can be explicitly checked in our previous paper\cite{BergvallPRB2013} in Eqs.~(B8)-(B9).}
This allows us to let $y_\alpha$ and $y^\prime_{\alpha^\prime}$ approach $-\infty$ in their respective lead,
with the added constraint that $y_{\alpha} < y^\prime_{\alpha^\prime} \rightarrow -\infty$.
This step properly identifies the $s$-matrix as connecting ideal reservoirs in the far distance
contacted to the scattering region through the ideal leads, in the spirit of the Landauer-B\"uttiker formalism.
If the coordinate systems are aligned as shown in Fig.~\ref{fig:coordinates},
we have to evaluate $\mathbf{G}_{\alpha \alpha^\prime}^{nn^\prime}(y,y^\prime;\hbar\omega)$
when $y\rightarrow +\infty$ and $y^\prime \rightarrow -\infty$. Bringing it all together we introduce
\begin{equation}
P^{n n^\prime}_{\alpha\alpha^\prime}(\hbar\omega) = 
\lim_{y\rightarrow\infty,y^\prime\rightarrow -\infty}
\vec{\phi}^{\dagger}_{\alpha n -}(\hbar\omega)
\mathbf{G}_{\alpha  \alpha^\prime}^{n n^\prime}(y,y^\prime;\hbar\omega) 
\vec{\phi}_{\alpha^\prime n^\prime +}(\hbar\omega).
\label{eq:projected_G}
\end{equation}
We now get the final expression for the $s$-matrix as
\begin{equation}
s^{m m^\prime}_{\alpha \alpha^\prime}(\hbar\omega) =
\frac{i}{\hbar}\sqrt{\frac{(\hbar v_f)^2 k_{\alpha m}(\hbar\omega)}{\hbar\omega-\Delta_\alpha}} \sqrt{\frac{(\hbar v_f)^2 k_{\alpha^\prime m^\prime}(\hbar\omega)}{\hbar\omega-\Delta_{\alpha^\prime}}}
P^{m m^\prime}_{\alpha \alpha^\prime}(\hbar\omega) 
- \vec{\phi}^\dagger_{\alpha m-}(\hbar\omega)\vec{\phi}_{\alpha m+}(\hbar\omega)\delta_{\alpha \alpha^\prime}^{m m^\prime}.
\label{eq:s-matrix}
\end{equation}
\newpage

\end{widetext}

In summary, the general strategy is to compute the propagator for the scattering part between leads,
then project as in Eq.~(\ref{eq:projected_G}) and use Eq.~(\ref{eq:s-matrix}) to get the $s$-matrix.
In the main text of this paper we use the analytic form for the propagator
for an AGNR with and without an impurity scattering center at $\rv_i$, obtained in Ref.~\onlinecite{BergvallPRB2013},
together with the above results to obtain analytic formulas for the conductance.

\end{document}